# HPL-GEM: Controlling High Pressure Laminates bulk resistivity with GEMs

**P.Vitulo**[a,b]

[a] *University of Pavia - Department of Physics,*
  *via Bassi, 6, 27100 Pavia, Italy*

[b] *INFN Pavia,*
  *via Bassi, 6, 27100 Pavia, Italy*

  *E-mail*: paolo.vitulo@unipv.it

ABSTRACT: We succeeded in modifying and controlling the electrical resistance of a standard High Pressure Laminate (HPL) panel through the use of a Gas Electron Multiplier (GEM) foil that has been embedded into the bulk of the HPL plate itself. Electrical characterizations were made and preliminary data show that this HPL-GEM embedded system can easily vary its bulk resistance by more than one order of magnitude. Data show that the bulk resistance change is exponential with the applied voltage to the embedded GEM.

**Contents**

1. **Introduction**

HPL is a standard material used in the furniture industry obtained by pressing and heating a pile of Kraft paper foils previously impregnated into a phenolic resin bath. Usually an overlay is added on both the external surface to give the product a better finish. The bulk electrical resistivity is usually not important to the end-user of the standard HPL market. However, HPL has been also successful used as the electrode's material in the production of the so called "Bakelite" Resistive Plate Chambers (RPC) detectors [1]. Their bulk resistivity plays a role in the rate capability of the detector itself. In view of the use of RPCs in the ATLAS and CMS experiments at LHC [2][3][4][5] the production of RPCs with electrode's bulk resistivity in the $10^{10}\ \Omega\ cm$ range has been the goal of an intense R&D study for several research groups. Most of these succeeded in lowering the HPL bulk resistivity by carefully modifying the heating protocol.

This work used a GEM foil embedded into a standard HPL panel to modify and control its resistance. A GEM[6] foil is a $50\ um$ polymer foil with 5 um copper clad on both the surfaces. A regular pattern of holes is chemically etched along the surface in a controlled and uniform way obtaining about 5000 holes/cm$^2$. The holes are about $70\ um$ in diameter and have $140\ um$ pitch. Application of a voltage bias (300 to 500 V) across the foil generates an intense electric field inside the holes that can be used in gaseous detectors as an electron multiplier. Our intention is to use a GEM embedded into the bulk of a HPL panel to increase the mobility or the number of the ionic carriers inside the material itself so as to increase the overall conductivity and in turn to decrease its resistivity. If this is the case, by tuning the bias across the GEM a modulation of the bulk resistivity would be readily obtained.
Preliminary electrical characterizations were made and show that this HPL-GEM embedded system can easily vary the bulk resistivity of the HPL-GEM system by more than an order of magnitude. Data show that the bulk resistivity change is exponential with the applied bias to the embedded GEM.

2. **The HPL-GEM**

Overall the system was obtained by starting from a pile of Kraft paper foils impregnated of phenolic resin[1] (PR) in the middle of which a standard 10 x10 cm$^2$ active area GEM foil has been embedded[2]. Modified HPL panels of 15 x15 cm$^2$ size area and 2 mm in thickness were so produced. The area of the panels was chosen to easy cover the active area of a GEM foil while the thickness was chosen as a remind of the panels used in RPC detector production.

---
[1] From Puricelli srl
[2] The work has been done at the CERN GEM workshop by R. De Oliveira

Figure 1 shows a result of such an embedded system for two different HPL laminates fabricated with different type of PR impregnated paper. The different pattern on the HPL's surfaces depends on the resin content of the paper foils forming the bulk of the HPL itself and on its spatial distribution, while the clear white edges around are probably due to the evaporation of the resin itself during the heating process used to obtain the laminates. We stress the fact that each GEM foil has not been trivially sandwiched between two existing HPL panels, but it has been put in the middle of a certain number of PR impregnated paper foils and processed together (heated and pressed) to form a whole system appearing as an HPL panel with a GEM foil embedded in the middle. Each GEM hence belongs to the bulk of an HPL panel. The GEM contacts were left outside to allow the bias to be applied to the GEM itself.

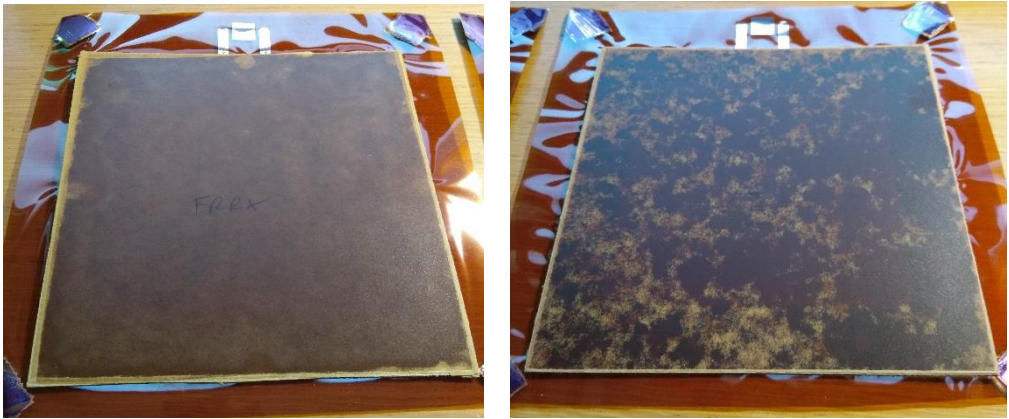

**Figure 1.** Two examples of HPL-GEM system where a GEM foil is embedded into the bulk of a HPL laminate. Visual differences of the laminates surfaces are due to the different type of phenolic resin impregnated foils used for their production. On the top of the pictures the copper contacts of the embedded GEMs are visible.

The idea of filling the bulk of Plastics or HPL with conductive materials of different structure (metallic planes, carbon black, metallic ions, graphite…) is certainly not new [7][8] but in this case the filler (the GEM) is an electrically active system. In the following it will be shown that this configuration has made it possible not only to modify the resistivity of the HPL, but also to control it by varying the applied voltage to the GEM.

3. **Measurements setup**

To measure the resistivity of a HPL-GEM system a fixed voltage (50 V or 100 V) is applied on the external surfaces of the HPL panel and the bulk current is monitored while a set of different bias values is applied to the GEM (20 V to 200 V). Copper electrodes were applied to the external surfaces of the HPL panel as shown in Figure 2, where also a guard ring is visible on one side of the HPL. A cross section sketch (not to scale) is also shown.

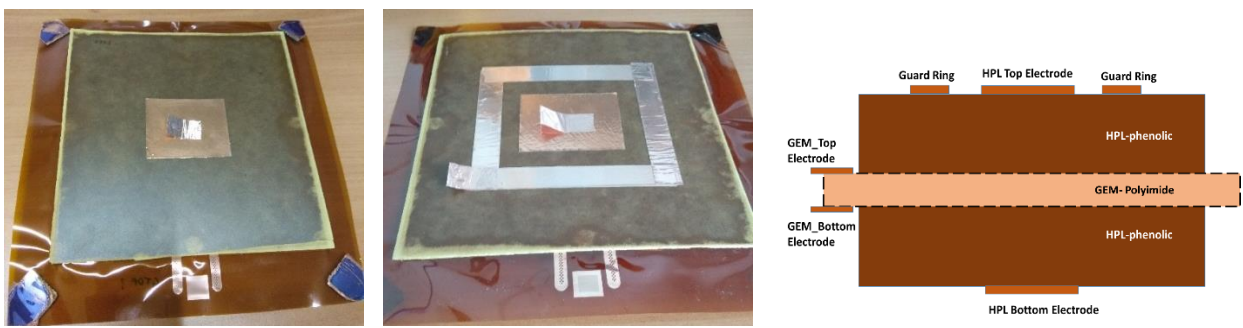

**Figure 2.** Copper electrodes (20 cm$^2$ size area) were applied on the external surfaces of the HPL-GEM system. Left: back side of the HPL-GEM. Center: top side of the HPL-GEM where also a guard ring is visible. GEM electrodes are visible

as copper strips on the bottom of the pictures. Right: A not to scale cross-section sketch of the HPL-GEM system with the external copper electrodes applied for the measurements.

Different High Voltage configurations have been considered so far for the powering scheme:

- NOHPLBIAS: only the GEM is powered and the bulk current across the HPL is measured
- NOGEMBIAS: only the HPL is powered and the current (and voltage) of the GEM is measured
- GEMRBIAS: GEM Reverse Bias, where the HPL is positively polarized while the GEM is negatively polarized
- GEMDBIAS: GEM Direct Bias, where both the HPL and the GEM are positively polarized.

Obviously one can reverse the sign of the polarization on the HPL (negative polarized as an example) and the sign of the GEM one accordingly, to respect the scheme of the configurations. The HPL and the GEM polarization were set independently by a Keithley 8467 Pico ammeter and a CAEN R1470ET 4 channel reversible HV Power Supply respectively. The Pico ammeter was also used to measure the bulk current across the HPL. The two instruments had floating grounds. Another possibility would be to use a divider to partition the HV from the top of the HPL to the GEM and down to the bottom of the HPL. However, in this work this last configuration was not adopted.

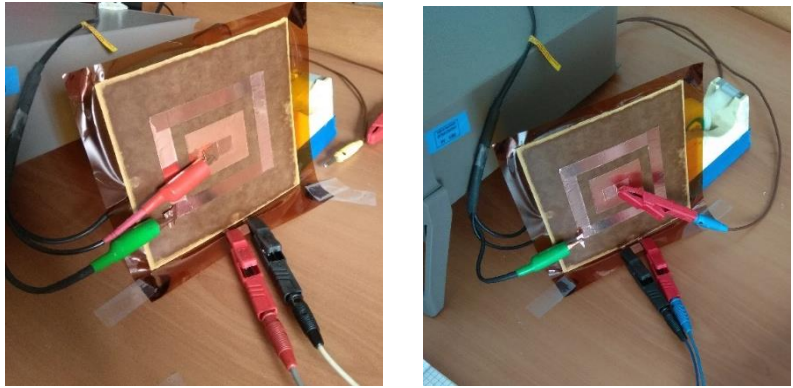

**Figure 3.** Examples of measurement configurations. Left: NOHPLBIAS, the HPL current is monitored through the Pico ammeter (using a triaxial cable connected to the two HPL external electrodes and also to the guard ring) while the GEM is independently powered through the CAEN power supply. Right: GEMDBIAS, both HPL and GEM are powered. In this case current or resistance of the HPL can be monitored through the Keithley Pico ammeter and the GEM current through the CAEN Power Supply.

In Figure 3 two of the configurations (NOHPLBIAS and GEMDBIAS) are shown. In the first case (left panel) the Pico ammeter was used only to measure the HPL bulk current (with a triaxial cable using also the guard ring) while the GEM was independently powered. The right picture shows the configuration in which both HPL and GEM were powered (the Keithley 8467 Pico ammeter can also provide High Voltage up to +/- 500 V). In this case current or resistance of the HPL is monitored through the Keithley Pico ammeter and the GEM current through the CAEN Power Supply.

### 4. Discussion and Results

As it was detailed before, the HPL-GEM system was produced by inserting a GEM foil into the standard impregnated paper foil packing used for the High Pressure Laminates and then heated and pressed. Some concerns should raise about the compression strength of the polyimide of which the GEM is made of, or about its melting point, but the pressure and temperature typical of the HPL production process ($< 2\ kPsi, < 200\ °C$) are negligible if compared to the standard values declared in the data sheets of typical polyimides as those used into the standard GEM foils[3]. Also a concern about a possible electrical short between the top and the bottom copper planes of the GEM foil may be shared. However, no shorts have been found on the panels we have tested. On the other hand, we measured a steady state resistance of few $M\Omega$ across the GEM foils of the HPL-GEM samples. This resistance is very small compared to that typical of a standard GEM foil (order of hundreds of $G\Omega$). As already pointed out we tend to exclude a change in the mechanical characteristic of the polyimide; moreover, if such a value of resistance was completely ascribed to a modified GEM thickness, this would be of the order of a $um$ (considering an average bulk resistivity of the polyimide of $10^{16}\ \Omega\ cm$). This would correspond to a compression factor of 50, which appears too high for the used temperature and pressure during the HPL-GEM system production. Instead we do not exclude the hypothesis of percolation of the resin through the GEM holes. The measured GEM's resistance would be the parallel of two resistances: that of the GEM's polyimide and that of the PR inside the holes. In this case and if all the GEM holes had been filled (holes represent about 20% of the GEM active area) the total resistance would be of the order of dozens of $M\Omega$ (considering a typical PR bulk resistivity of $10^{11}\ \Omega\ cm$). This outcome is considered a good indication since usually the type of PRs and relative additives are hardly known. We also measured the capacitance[4] of one HPL-GEM as a whole and by considering it as a series of 3 capacitors (see the sketch of Figure 2) : the capacitor formed by one external HPL electrode and one (copper) face of the GEM, the GEM itself, and finally, the capacitor formed by the other (copper) face of the GEM and the other HPL external electrode. The specific capacitance of the embedded GEM was measured to be $Cp = 116\ pF/cm^2$ with a parallel specific resistance $Rp = 14.3\ k\Omega/cm^2$ at a frequency of $20\ Hz$ while the specific capacitance of a similar GEM ($10 \times 10\ cm^2$) in air is about $60\ pF/cm^2$.

### 4.1 A simple electrical model

From the above measurements a simple electrical model of the tested HPL-GEM system is derived by considering HPL and GEM as lumped circuits made out of parallel connections of capacitances and resistors. The model with the measured capacitance and resistance at 20 Hz is shown in Figure 4. Also the surface HPL resistance is considered although not yet measured in this test.

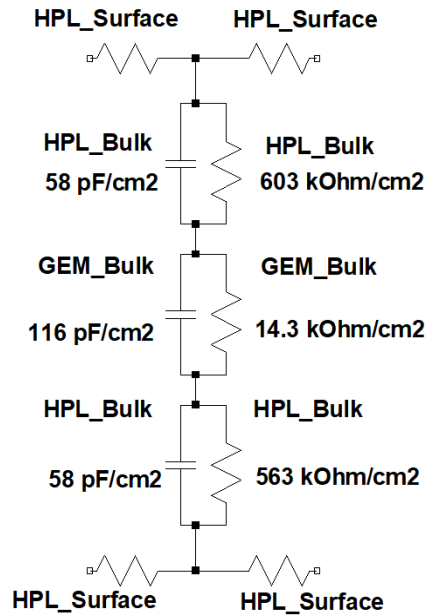

**Figure 4.** Simple electrical model of the tested HPL-GEM system. The values correspond to the measured capacitance and resistance at 20 Hz. Measurements were done with a LCR at a frequency of 20 Hz and with a bias of 2 V.

---

[3] See for example Apical-Kaneka Data Sheet or DuPont Kapton HN Data Sheet
[4] For the capacitance measurements we used an Agilent 4284A precision LCR Meter (20Hz-1MHz)

### 4.2 NOGEMBIAS Configuration Results

We observed a charging up at the GEM interface while powering the HPL only in the NOGEMBIAS configuration (see Sec. 3 for the conventions). The GEM potential was measured by a tester while the HPL was powered at different voltages up to 200 V, as shown in Figure 5 (left panel). In the plot a second order polynomial function has also been drawn to guide the eyes. In the same Figure 5 (right panel) the corresponding HPL characteristic curve (Voltage vs Current, black circles) is also shown. Each point corresponds to a measurement duration of 1 minute in which the HPL current was acquired by the Pico ammeter (with an integration time of about 1 s) and its current value is the average of the last three acquisition points. This behavior is typical of a space charge limited conduction regime (in which the current density is proportional to the square of the applied voltage) [9] [10]. A power function fit is superimposed as a continuous line. The black triangles points refer to another HPL-GEM prototype made with the same type of impregnated paper foils. Also in this case a power function fit is superimposed as a dashed line.

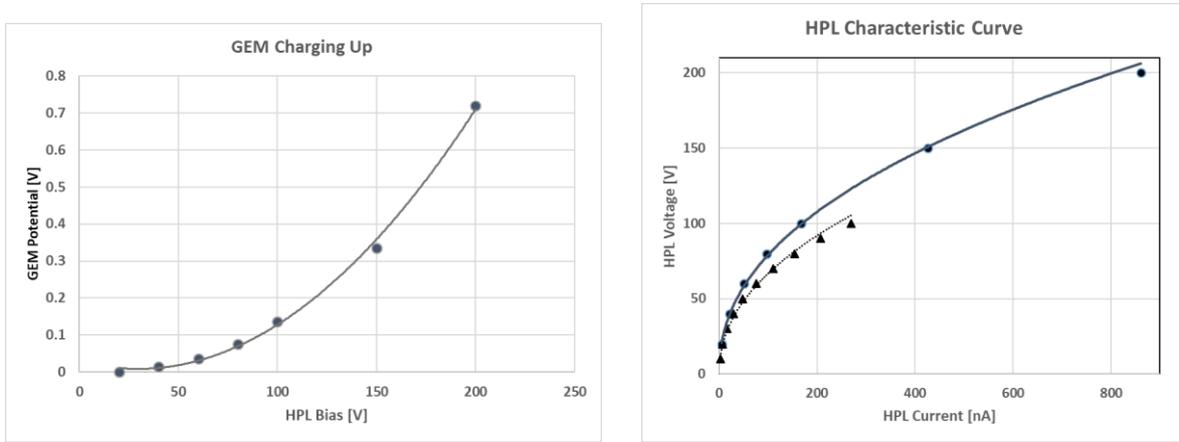

**Figure 5.** NOGEMBIAS configuration. Left panel: GEM Potential as measured by a tester during the application of the bias to the HPL. The GEM shows a charging up effect during the test. A second order polynomial has been drawn to guide the eyes. Right panel: HPL characteristic curve (Voltage vs Current, black circles) as measured by the Keithley Pico ammeter that shows a current proportional to the square of the applied voltage. A power function fit is superimposed as continuous line. The black triangles points refer to another HPL-GEM prototype made with the same type of impregnated paper foils. Also in this case a power function fit is superimposed as a dashed line.

### 4.3 GEMRBIAS Configuration Results

In Figure 6, measurements on one HPL-GEM are shown in the GEMRBIAS configuration, where +50 V bias was applied to the HPL and the GEM bias was varied up to -200 V. Also in this test each point corresponds to 1-minute measurement during which also the resistance of the HPL was monitored. Moreover, the initial and the final current of the GEM was acquired. Left panel of Figure 6 shows the GEM characteristic curve (Voltage vs Current) in which this initial GEM current has been used. A straight line is superimposed that show the ohmic behavior of the GEM. In the right panel of Figure 6 the HPL resistance (in $G\Omega$) is plotted (black circles) as a function of the GEM bias showing the effectiveness of the GEM in lowering the HPL resistance by more than one order of magnitude (a factor 33 in this case). An exponential curve plus a constant term has been superimposed to the plot to guide the eyes. The red triangles marks refer to another HPL-GEM prototype made with the same type of impregnated paper foils tested in the same configuration (GEMRBIAS, HPL voltage fixed at +50V). Also in this case an order of magnitude reduction of the HPL resistance has been reached.

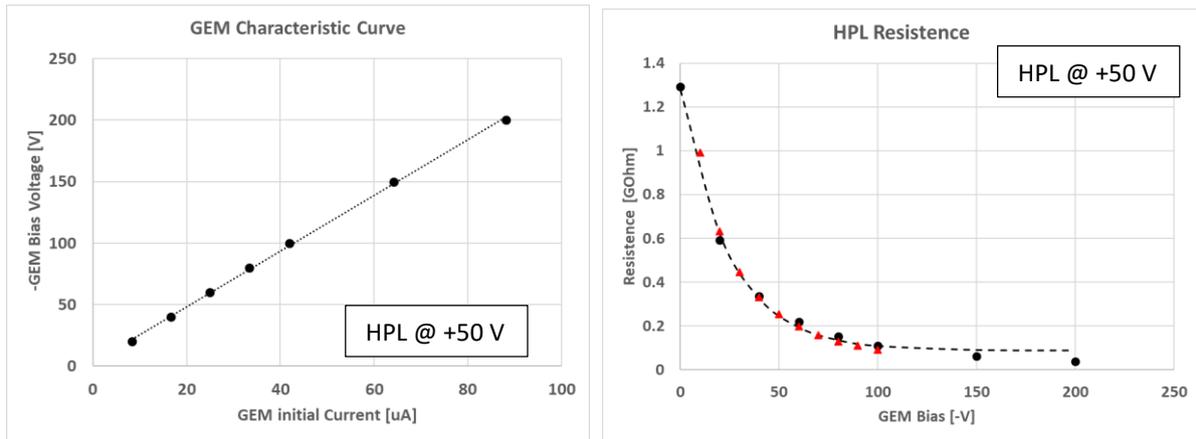

**Figure 6.** HPL-GEM System in the GEMRBIAS configuration. Left panel: (V,I) Characteristic curve of the embedded GEM while the HPL was set at a fixed voltage of +50 V. GEM current are in uA. The GEM initial current (as explained in the text) has been used. A straight line has been added that shows the Ohmic behavior of the embedded GEM. Right panel (black circles marks): HPL resistance of the HPL (kept at +50 V) as a function of the GEM bias. A resistance decrease of more than one order of magnitude has been reached in this test. An exponential curve plus a constant term has been superimposed to guide the eyes. Red triangles marks refer to another HPL-GEM prototype made with the same type of impregnated paper foils tested in the same configurations.

### 4.4 NOHPLBIAS Configuration Results

We stress that the GEM insertion into the HPL panel has an active role in modifying and controlling the resistance (and hence the resistivity) of the HPL itself. In fact, by reversing the polarity of the GEM bias with respect to the HPL one (GEMDBIAS configuration) is it possible to increase the HPL resistance by at least the same factor as before. A hint of this behavior can be obtained in the NOHPLBIAS configuration in which we applied only the bias to the GEM and the corresponding HPL density current was measured. Figure 7, which represents the measured HPL current density (in $nA/cm^2$) as a function of the GEM bias (in Volt) shows this behavior.

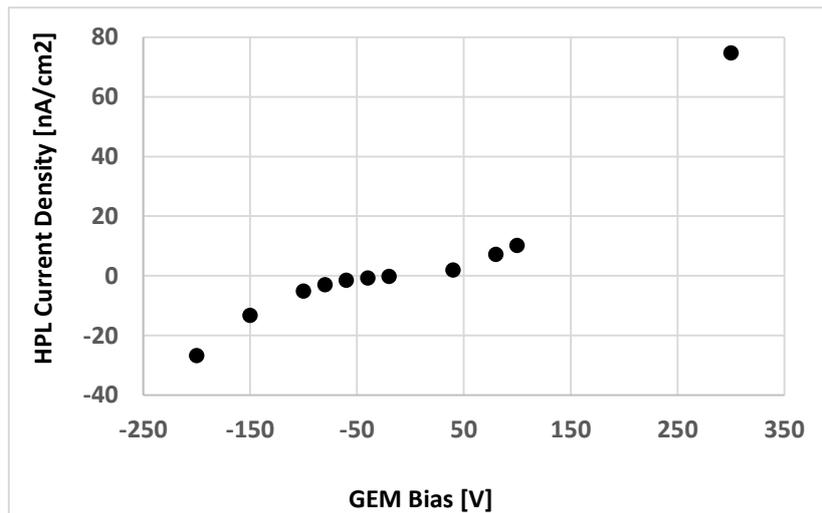

**Figure 7.** HPL-GEM System in the NOHPLBIAS configuration. In this setup, voltage was applied only to the GEM and the HPL current measured through the Pico ammeter. The marks show the HPL density current as a function of the GEM bias.

### 4.5 GEMDBIAS Configuration Results

As already pointed out, the GEM can be polarized with the same polarity as the HPL. The effect of this configuration is to lower the conduction of the HPL through the effect of a voltage barrier that halt the injection of charge from the HPL electrodes and hence increases its resistance. This is true as long as the GEM bias is lower than the HPL bias. The experimental points of Figure 8 (left panel) show this behavior when the resistance of the HPL is plotted as a function of the GEM bias. For a GEM bias higher than the HPL one (in this case 50 Volt) then the effect is again a reduction of the resistance. This behavior is equivalent with a shift of the curve of Figure 7 to the right by 50 Volts provided that the resistance of the HPL is set positive also for a negative value of the HPL current. That the GEM has an important role in the overall electrical conduction of the system can be shown by considering the difference in the GEM current after each 1-minute acquisition time compared to the measured HPL density current (right panel of Figure 8). It shows that the higher is the difference in the GEM current the higher will be the HPL current density (and hence the lower its resistivity).

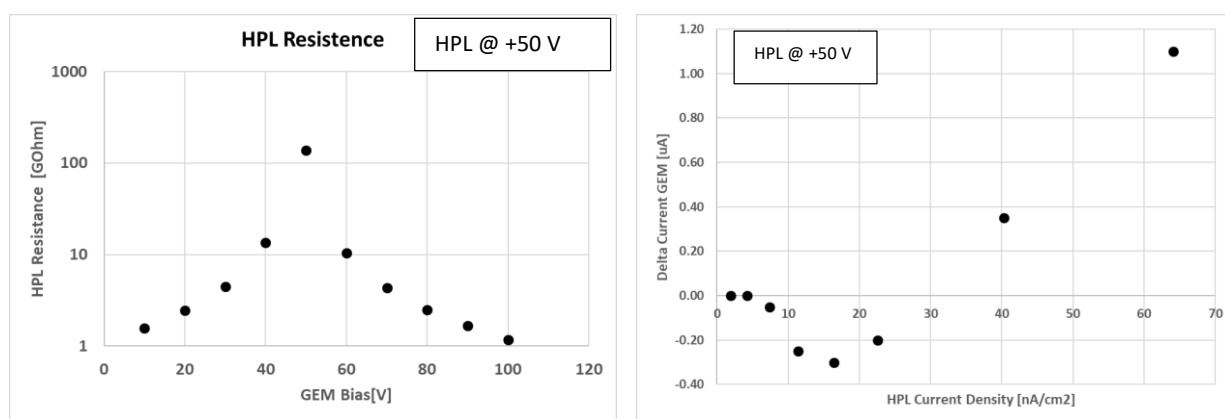

**Figure 8.** HPL-GEM System in the GEMDBIAS configuration. Left: in this setup, GEM and HPL has the same polarity and for this test the HPL was kept at +50 Volts. The marks show the measured HPL resistance as a function of the GEM bias. The HPL resistance is considered positive also for a negative current of the HPL. Right: difference of GEM current (after each 1-minute acquisition) as a function of the HPL current density.

Also in all these tests each point corresponds to a 1-minute measurement in which the HPL current/resistance has been measured (with about 1 s integration time) the value being the average of the last 3 points of the acquisition. The HPL resistance reduction/increase factor effect for this configuration is higher with respect to that seen in Figure 6. All these effects have to be studied in detail according to the conduction mechanisms involved when the GEM is operative in this HPL-GEM system and will be the subject of a future work.

### 5. Conclusions and Outlooks

We have presented the results of preliminary tests done on a new system called HPL-GEM in which two technologies have been exploited: a Gas Electron Multiplier foil has been embedded into the bulk of High Pressure Laminate panel like that used as resistive electrodes for the Resistive Plate Chambers. The basic idea was to act on the HPL system overall resistivity and to control it. The data show that the resistivity of the HPL panel can be modified and controlled by varying the voltage applied to the GEM. The resistivity variation is more than one order of magnitude. The conduction mechanisms through which this is possible are still under study and a lot of work has to be done as concerning the HPL materials and the geometries (the numbers of GEM foils that can be embedded related to the HPL panel thickness). It is then easy to predict that also the surface resistivity of a HPL-GEM system could be changed and tests are ongoing on this subject. Also at level of construction more data have to be analyzed through the production of different prototypes. Nevertheless, an important application would be the use of this HPL-GEM system as the actual electrodes of a RPC detector where their resistivity plays an important role; in this case also the transparency of the signal need to be addressed.